\documentclass[11pt,a4paper,twoside]{article}  %
\usepackage[american]{babel}     
\usepackage{amsmath,amsthm,amssymb,latexsym}
\usepackage{amsfonts}           
\usepackage{geometry}
\usepackage{graphicx}         
\usepackage{fancyhdr}
\usepackage{tikz}
\usepackage{color}

\def\RR{\mathbb{R}}

\newtheorem{definition}{Definition}[section]

\newtheorem{lemma}{Lemma}[section]   
\newtheorem{pro}{Proposition}[section]
\newtheorem{cor}{Corollary}[section]    
\newtheorem{remark}{Remark}[section]   
\newtheorem{conj}{Conjecture}[section]

\newenvironment { abstract }

\title{ \bf Quantifying Networks Complexity from Information Geometry Viewpoint }

\author{Domenico Felice\\
School of Science and Technology \\University of Camerino, I-62032 Camerino, Italy\\
INFN-Sezione di Perugia, Via A. Pascoli, I-06123 Perugia, Italy
\and 
Stefano Mancini\\
School of Science and Technology \\University of Camerino, I-62032 Camerino, Italy\\
INFN-Sezione di Perugia, Via A. Pascoli, I-06123 Perugia, Italy
\and
Marco Pettini\\
Centre de Physique Th\'eorique, UMR7332, and Aix-Marseille University, \\ Luminy Case 907, 13288 Marseille, France}

\begin{document}

\maketitle

\begin{abstract}
We consider a Gaussian statistical model whose parameter space is given by the variances of random variables.
Underlying this model we identify networks by interpreting random variables as sitting on vertices and their correlations as weighted edges among vertices.
We then associate to the parameter space a statistical manifold endowed with a Riemannian metric structure (that of Fisher-Rao). Going on, in analogy with the
microcanonical definition of entropy in Statistical Mechanics, we 
introduce an entropic measure of networks complexity.
We prove that it is invariant under networks isomorphism.
Above all, considering networks as simplicial complexes, we evaluate this entropy on simplexes and find that it monotonically increases with their dimension. 
\end{abstract}

\date{}


\section{Introduction}

The notion of complexity is central in many branches of science. Common understanding tells us what is simple and complex, however formalizing this rather elusive notion results a daunting task \cite{GM95}. That has led to a variety of definitions and measures of complexity.  
Among them one can consider the statistical ones \cite{FC98} with an example provided by the Fisher-Shannon information \cite{LR12}.

The notion of complexity is also relevant when dealing with networks. Actually complex networks have become one of the prominent tools in the study of social, technological and biological science \cite{CNet1}. In particular, the statistical approach to complex networks is a dominant paradigm in describing natural and societal systems \cite{CNet2}. 
By means of statistical complexity there is the possibility to consider both information about and structure of networks  \cite{LMC95}.

Information geometry concerns the possibility of dealing with statistical models by using differential geometry tools \cite{AN00}. This is realized by analyzing the spaces of probability distributions as Riemannian differentiable manifolds.   
Information geometry has been already used to study the complexity of informational geodesic flows on curved statistical manifolds \cite{C09,ACK10} and to formalize the idea that in a complex system the whole is more than the sum of its parts \cite{Jost011}.

Here, we resort to information geometry to introduce a statistical measure of networks complexity. 
We start considering a statistical model with underlying network by interpreting random variables as sitting on vertices and their correlations as weighted edges among vertices. 
Specifically, from Section \ref{section2} on, we shall focus on Gaussian statistical models, motivated by the fact that very often in real world random variables are Gaussian distributed (with parameter space given by the variances of random variables).
For the sake of simplicity we shall consider presence/absence of correlations thus taking weights of edges simply equal to 1 or 0.
Since Gaussian probability distributions are parametrized by real-valued variables it is possible to provide a $C^{\infty}$-differentiable structure upon this set. In this way, we are able to consider a differentiable statistical manifold \cite{AN00}. The information of the system underlying the manifold is provided by the Fisher information matrix which also provides a Riemannian metric to the parameter space. The reason of this choice is that the properties of a given network, that is a discrete object, are lifted to the geometric structure of a manifold, that is a differentiable object.

Then, in Section \ref{section3} we introduce a measure of complexity related to the volume of the manifold. This is inspired by the microcanonical definition of thermodynamical entropy in  Statistical Mechanics as the logarithm of the phase space volume.
After dealing with the difficulty of defining a proper volume on a manifold that results non compact, we show that the introduced measure of complexity is invariant under networks isomorphism.

Finally, in Section \ref{section4} we consider networks as simplicial complexes, we evaluate the measure of complexity on simplexes and show that it increases with their dimension. This reveals its sensitivity to topological network features, in contrast for example to the Fisher-Shannon information which does not distinguish among Gaussian models.


\section{Gaussian statistical manifolds and networks}\label{section2}

We start considering a set of $n$ random variables $x_1,\ldots,x_n$ defined on the continuous real alphabet with 
a joint probability distribution $p$, a function $p:\RR^n\rightarrow\RR$ satisfying the conditions
\begin{equation*}
p(x)\geq 0\; (\forall x\in \RR^n)\quad \mbox{and}\quad \int_{\RR^n} dx \;p(x)=1.
\end{equation*}
Next we consider  a family $\cal P$ of such probability distributions parametrized by $m$ real-valued variables 
$(\theta ^1,\ldots\theta^m)$ so that
\begin{equation*}
{\cal P}=\{p_\theta=p(x;\theta)|\theta=(\theta^1,\ldots,\theta^m)\in\Theta\},
\end{equation*}
where $\Theta\subseteq\RR^m$ and the mapping $\theta\rightarrow p_\theta$ is injective. 
Intended in such a way $\cal P$ is an $m$-dimensional statistical model on $\RR^n$.

The mapping $\varphi:{\cal P}\rightarrow \RR^m$ defined by $\varphi(p_\theta)=\theta$ allows us to consider $\varphi=[\theta^i]$ as a coordinate system for $\cal P$. Assuming parametrizations which are $C^\infty$ we can turn 
$\cal P$ into a $C^\infty$ differentiable manifold ($\cal P$ is thus called statistical manifold) \cite{AN00}.

Let ${\cal P}=\{p_\theta|\theta\in\Theta\}$ be an $m$-dimensional statistical model. Given a point $\theta$, the Fisher information matrix of $\cal P$ in $\theta$ is the $m\times m$ matrix $G(\theta)=[g_{\mu\nu}]$, where the $\mu,\nu$ entry is defined by
\begin{equation}
g_{\mu\nu}(\theta):=\int_{\RR^n} dx p(x\arrowvert\theta)\partial_\mu\log p(x\arrowvert\theta)\partial_\nu\log p(x\arrowvert\theta),
\label{gFR}
\end{equation}
with $\partial_\mu$ standing for $\frac{\partial}{\partial\theta^\mu}$. The matrix $G(\theta)$ is symmetric, positive semidefinite and determines a Riemannian metric on the parameter space $\Theta$ \cite{AN00}.

From here on we assume to deal with an $n$-variate Gaussian probability distribution 
for the $n$ random variables, i.e. 
\begin{equation}
p(x\arrowvert\theta)=\frac{1}{\sqrt{(2\pi)^n\det C}}\exp\left[-\frac 1 2 x^t {C}^{-1}x\right],
\label{PxT}
\end{equation}
where $C$ denotes the covariance matrix and $t$ the transposition. Furthermore, assume that the parameters are the variances
of the random variables
\begin{equation*}
\theta^i= \int_{\RR^n} dx\; p(x|\theta) x_i^2, \quad i=1,\ldots,m=n,
\end{equation*}
while it is assumed that the random variables have zero mean.
Generally speaking the parameters can be regarded as the pieces of information about the system 
(random variables) one can access.

Then the statistical manifold is determined by
\begin{equation}\label{parameterspace}
\Theta=\{\theta\in \RR^n|C(\theta)>0 \}.
\end{equation}
At this point we can interpret random variables as sitting on vertices of a network and correlations of random variables  as weighted edges among vertices of such a network. For the sake of simplicity we shall consider the weights, i.e. 
 the non-diagonal entry $c_{ij}$ of the covariance matrix $C$, to be either $1$ or $0$. 

Given the formal definition of the Fisher-Rao metric tensor (\ref{gFR}), in order to make it of practical use it is crucial to try to work out a simple and more explicit analytical relation between the entries of the matrix $G$ and those of the covariance matrix $C$. It turns out that such a simple relation actually exists. Let us see how things proceed.

Because of Eq.\eqref{PxT} we note that Eq.\eqref{gFR} involves a Gaussian Integral. However, before evaluating it, let us study the function
\begin{equation}
\label{fmunu}
f_{\mu\nu}(x):=\partial_\mu\log p(x\arrowvert\theta)\partial_\nu\log p(x\arrowvert\theta).
\end{equation}
By means of logarithm's properties we can write
\begin{equation}\label{log}
\log[p(x|\theta)]=-\frac 1 2\Bigg[\log[(2\pi)^n\det C(\theta)]+\sum_{\alpha,\beta=1}^n c_{\alpha\beta}^{-1}(\theta)x_\alpha x_\beta\Bigg],
\end{equation}
where $c_{\alpha\beta}^{-1}$ is the entry $\alpha\beta$ of the inverse of the covariance matrix $C$. Then the derivative $\partial_\mu$ of Eq.\eqref{log} reads
\begin{equation*}
\partial_\mu \log[p(x|\theta)]=-\frac 1 2\Bigg[\frac{\partial_\mu(\det C)}{\det C}+\sum_{\alpha,\beta=1}^n \partial_\mu(c_{\alpha\beta}^{-1})x_\alpha x_\beta\Bigg]. 
\end{equation*}
Recall that the following relations hold
$$\partial_\mu(\det C(\theta))=\det C(\theta) \mbox{Tr}(C(\theta)\,\partial_\mu(C(\theta)));$$
$$\partial_\mu(C(\theta))=\Big[\frac{\partial c_{ij}}{\partial\theta^\mu}\Big]_{ij}.$$

Hence, observing that the only non zero entries of the matrix $\partial_\mu(C(\theta))$ are $\frac{\partial c_{\mu\mu}}{\partial\theta^\mu}=1$, we find
\begin{equation}
\partial_\mu\det (C(\theta))=c_{\mu\mu}^{-1}(\theta)\det (C(\theta)).
\label{derivativeofdet}
\end{equation}

Furthermore, for any invertible matrix $A$ it is well-known the following relation about the derivative of the inverse matrix \cite{Cor09}
\begin{equation*}
\partial_\mu (A^{-1}(\theta))=-A^{-1}\big(\partial_\mu(A)\big)A^{-1}.
\end{equation*}
Then the derivative of the inverse of the covariance matrix $C$ reads $
\partial_\mu (C^{-1}(\theta))=-\Big[c_{i\mu}^{-1}(\theta)c_{j\mu}^{-1}(\theta)\Big]_{ij},
$ and so the entry $ij$ is given by the relation
\begin{equation}
\partial_\mu(c_{ij}^{-1})(\theta)=-c_{i\mu}^{-1}(\theta)c_{j\mu}^{-1}(\theta).
\label{derivativeinverse}
\end{equation}
Thanks to Eqs.\eqref{derivativeofdet} and \eqref{derivativeinverse} we obtain
\begin{equation}
\partial_\mu \log[p(x|\theta)]=-\frac 1 2\Bigg[c_{\mu\mu}^{-1}-\sum_{\alpha,\beta=1}^n c_{\alpha\mu}^{-1}c_{\beta\mu}^{-1}x_\alpha x_\beta\Bigg].
\label{derivativelog}
\end{equation}

We are now going to evaluate the Gaussian integrals in Eq.\eqref{gFR}. 
Using Eq.\eqref{derivativelog}, the function $f_{\mu\nu}(x)$ of Eq.\eqref{fmunu} reads
\begin{equation}
f_{\mu\nu}(x)=\frac 1 4\Bigg[c_{\mu\mu}^{-1}-\sum_{\alpha,\beta=1}^n c_{\alpha\mu}^{-1}c_{\beta\mu}^{-1}x_\alpha x_\beta\Bigg] \Bigg[c_{\nu\nu}^{-1}-\sum_{\alpha,\beta=1}^n c_{\alpha\nu}^{-1}c_{\beta\nu}^{-1}x_\alpha x_\beta\Bigg].
\label{f}
\end{equation}

For a differentiable function $f(x)$ and a symmetric definite-positive  $n\times n$ matrix
$A=(a_{ij})$ it results
\begin{equation}
\int dx f(x)\exp\Big[-\frac{1}{2}\sum_{i,j=1}^na_{ij}x_ix_j\Big]
=\sqrt{\frac{(2\pi)^n}{\det A}}\exp\left[\frac 1 2 \sum_{i,j=1}^n 
a_{ij}^{-1}\frac{\partial}{\partial x_i}\frac{\partial}{\partial x_j}\right]f |_{x=0},
\label{IG}
\end{equation}
where $a_{ij}^{-1}$ is the entry $ij$ of the inverse of the matrix $A$ and the exponential means the power series over its argument (the differential operator). 

Substituting expression \eqref{PxT} into the relation \eqref{gFR} and employing Eq.\eqref{IG} we find
\begin{eqnarray}
g_{\mu\nu}=&&\frac{1}{\sqrt{(2\pi)^n\det C}}\int dx f_{\mu\nu}(x) \exp\left[-\frac 1 2 x^t C^{-1} x\right]\nonumber\\
&&=\exp\left[\frac 1 2 \sum_{i,j=1}^n 
c_{ij}\frac{\partial}{\partial x_i}\frac{\partial}{\partial x_j}\right]f_{\mu\nu} |_{x=0}.
\label{Gint}
\end{eqnarray}

\begin{lemma}\label{lemma1}
Let us set $D:=\frac 1 2 \sum_{i,j=1}^nc_{ij}\frac{\partial}{\partial x_i}\frac{\partial}{\partial x_i}$; expanding the right-hand side of \eqref{Gint} we have that
\begin{equation}
g_{\mu\nu}(\theta)=f_{\mu\nu}(0)+Df_{\mu\nu} |_{x=0}+\frac 1 2 D^2f_{\mu\nu} |_{x=0},
\label{gexp}
\end{equation}
with
\begin{equation}
Df_{\mu\nu}|_{x=0}=
-\frac 1 4\sum_{i,j=1}^nc_{ij}(c_{i\mu}^{-1}c_{j\mu}^{-1}c_{\nu\nu}^{-1}+c_{i\nu}^{-1}c_{j\nu}^{-1}c_{\mu\mu}^{-1}),
\label{D}
\end{equation}
and
\begin{eqnarray}
\frac 1 2 D^2f_{\mu\nu}|_{x=0}&=&\frac 1 8 \Bigg(\sum_{i,j,h,k=1}^n c_{ij}c_{hk}\frac{\partial}{\partial x_i}\frac{\partial}{\partial x_j}\frac{\partial}{\partial x_h}\frac{\partial}{\partial x_k}\Bigg)f_{\mu\nu}|_{x=0}\nonumber\\
&=&\frac 1 8\sum_{i,j,h,k=1}^n c_{ij}c_{hk}\Big(c_{i\mu}^{-1}c_{j\mu}^{-1}c_{h\nu}^{-1}c_{k\nu}^{-1}+c_{k\mu}^{-1}c_{j\mu}^{-1}c_{h\nu}^{-1}c_{i\nu}^{-1}\nonumber\\
&&+c_{h\mu}^{-1}c_{j\mu}^{-1}c_{k\nu}^{-1}c_{i\nu}^{-1}+c_{k\mu}^{-1}c_{i\mu}^{-1}c_{h\nu}^{-1}c_{j\nu}^{-1}+c_{h\mu}^{-1}c_{i\mu}^{-1}c_{k\nu}^{-1}c_{j\nu}^{-1}\nonumber\\
&&+c_{i\nu}^{-1}c_{j\nu}^{-1}c_{h\mu}^{-1}c_{k\mu}^{-1}\Big).
\label{D2}
\end{eqnarray}
\end{lemma}

{\bf Proof.}\quad Letting $i,j\in\{1,\ldots,n\}$, by a straightforward calculation, we have

\begin{eqnarray*}
\frac{\partial}{\partial x_i}\Bigg(\frac{\partial f_{\mu\nu}}{\partial x_j}\Bigg)(x)&=&-\frac 1 2 c_{i\mu}^{-1}c_{j\mu}^{-1}c_{\nu\nu}^{-1}+\frac 1 2c_{i\mu}^{-1}c_{j\mu}^{-1}\sum_{\alpha,\beta=1}^nc_{\alpha\nu}^{-1}c_{\beta\nu}^{-1}x_\alpha x_\beta\\
&&+\sum_{\alpha=1}^nc_{\alpha\mu}^{-1}c_{j\mu}^{-1}x_\alpha \;\sum_{\alpha=1}^nc_{\alpha\nu}^{-1}c_{i\nu}^{-1}x_\alpha+\sum_{\alpha=1}^nc_{\alpha\mu}^{-1}c_{i\mu}^{-1}x_\alpha \;\sum_{\alpha=1}^nc_{\alpha\nu}^{-1}c_{j\nu}^{-1}x_\alpha\\
&&-\frac 1 2 c_{i\nu}^{-1}c_{j\nu}^{-1}c_{\mu\mu}^{-1}+\frac 1 2 c_{i\nu}^{-1}c_{j\nu}^{-1}\sum_{\alpha,\beta=1}^nc_{\alpha\mu}^{-1}c_{\beta\mu}^{-1}x_\alpha x_\beta.
\end{eqnarray*}
Taking the sum over $i,j\in\{1,\ldots,n\}$ and evaluating the above expression for $x=0$, we obtain Eq.\eqref{D}.

Then, letting $i,j,h,k\in\{1,\ldots,n\}$, we have 
\begin{eqnarray*}
\frac{\partial}{\partial x_h}\Bigg(\frac{\partial}{\partial x_k}\frac{\partial}{\partial x_i}\frac{\partial }{\partial x_j}f_{\mu\nu}\Bigg)(x)&=&c_{i\mu}^{-1}c_{j\mu}^{-1}c_{h\nu}^{-1}c_{k\nu}^{-1}+c_{k\mu}^{-1}c_{j\mu}^{-1}c_{h\nu}^{-1}c_{i\nu}^{-1}+c_{h\mu}^{-1}c_{j\mu}^{-1}c_{k\nu}^{-1}c_{i\nu}^{-1}\nonumber\\
&&+c_{k\mu}^{-1}c_{i\mu}^{-1}c_{h\nu}^{-1}c_{j\nu}^{-1}+c_{h\mu}^{-1}c_{i\mu}^{-1}c_{k\nu}^{-1}c_{j\nu}^{-1}+c_{i\nu}^{-1}c_{j\nu}^{-1}c_{h\mu}^{-1}c_{k\mu}^{-1}.\nonumber\\
\label{derivata4}
\end{eqnarray*}
Taking the sum over $i,j,h,k\in\{1,\ldots,n\}$ Eq.\eqref{D2} straightforwardly follows.

Finally, thanks to the above expression we have that the expansion with respect to the variable $x\in\RR$ in the right-hand side of 
Eq.\eqref{Gint}, around $x=0$, only contains terms up to the second order. $\hfill\Box$

\bigskip

We are now ready to state the main result of this Section.

\begin{pro}
The entry $\mu\nu$ of the metric the tensor \eqref{gFR} results
\begin{equation}
g_{\mu\nu}=\frac 1 2 (c_{\mu\nu}^{-1})^{2},
\label{gvsC}
\end{equation}
where $c_{\mu\nu}^{-1}$ is the entry $\mu\nu$ of the inverse of the covariance matrix $C$.
\label{main}
\end{pro}

{\bf Proof.} From Lemma \ref{lemma1} we can write 
\begin{equation*}
g_{\mu\nu}=f_{\mu\nu}(0)+D(f_{\mu\nu})|_{x=0}+\frac{1}{2}D^2(f_{\mu\nu})|_{x=0}.
\end{equation*}
From Eq.\eqref{f} it follows that $f_{\mu\nu}(0)=\frac 1 4 c_{\mu\mu}^{-1}\;c_{\nu\nu}^{-1}$. Using Eqs.\eqref{D} and \eqref{D2} we have
\begin{eqnarray}
\label{g}
g_{\mu\nu}&=&\frac 1 4 c_{\mu\mu}^{-1}\;c_{\nu\nu}^{-1}-\frac 1 4\sum_{i,j=1}^nc_{ij}(c_{i\mu}^{-1}c_{j\mu}^{-1}c_{\nu\nu}^{-1}+c_{i\nu}^{-1}c_{j\nu}^{-1}c_{\mu\mu}^{-1})\nonumber\\
&&+\frac 1 8\sum_{i,j,h,k=1}^n c_{ij}c_{hk}\Big(c_{i\mu}^{-1}c_{j\mu}^{-1}c_{h\nu}^{-1}c_{k\nu}^{-1}+c_{k\mu}^{-1}c_{j\mu}^{-1}c_{h\nu}^{-1}c_{i\nu}^{-1}+c_{h\mu}^{-1}c_{j\mu}^{-1}c_{k\nu}^{-1}c_{i\nu}^{-1}\nonumber\\
&&+c_{k\mu}^{-1}c_{i\mu}^{-1}c_{h\nu}^{-1}c_{j\nu}^{-1}+c_{h\mu}^{-1}c_{i\mu}^{-1}c_{k\nu}^{-1}c_{j\nu}^{-1}+c_{i\nu}^{-1}c_{j\nu}^{-1}c_{h\mu}^{-1}c_{k\mu}^{-1}\Big).\\\nonumber
\end{eqnarray}
We now notice that 
\begin{equation*}
\sum_{i,j=1}^nc_{ij}c_{i\mu}^{-1}=\left\{\begin{array}{ll}
1&\mbox{if}\;j=\mu\\
0&\mbox{if}\;j\neq\mu
\end{array}\right.,
\end{equation*}
and
\begin{equation*}
\sum_{h,k=1}^nc_{hk}c_{h\mu}^{-1}=\left\{\begin{array}{ll}
1&\mbox{if}\;k=\mu\\
0&\mbox{if}\;k\neq\mu
\end{array}\right..
\end{equation*}
Hence, by using Eq.\eqref{g} and filling out we obtain
\begin{eqnarray}
g_{\mu\nu}&=&\frac 1 4 c_{\mu\mu}^{-1}\;c_{\nu\nu}^{-1}-\frac 1 4 c_{\mu\mu}^{-1}\;c_{\nu\nu}^{-1}-\frac 1 4 c_{\mu\mu}^{-1}\;c_{\nu\nu}^{-1}\nonumber\\
&&
+\frac 1 8 c_{\mu\mu}^{-1}\;c_{\nu\nu}^{-1}+\frac 1 8 c_{\nu\mu}^{-1}\;c_{\nu\mu}^{-1}+\frac 1 8 c_{\nu\mu}^{-1}\;c_{\nu\mu}^{-1}
\nonumber\\
&&+\frac 1 8 c_{\nu\mu}^{-1}\;c_{\nu\mu}^{-1}+\frac 1 8 c_{\nu\mu}^{-1}\;c_{\nu\mu}^{-1}+\frac 1 8 c_{\nu\nu}^{-1}\;c_{\mu\mu}^{-1}\nonumber\\
&=&\frac 1 2 (c_{\mu\nu}^{-1})^2. \nonumber
\end{eqnarray}
$\hfill\Box$


\section{A statistical measure of complexity}\label{section3}

Let us consider the $n$-dimensional Gaussian statistical model $p(x|\theta)$ of Eq.\eqref{PxT} with the parameter space $\Theta$ given by the Eq.\eqref{parameterspace}. Let $\mathcal{M}=(\Theta,g)$ be the statistical Riemannian manifold with metric tensor 
\begin{equation*}
g:=\sum_{\mu,\nu=1}^ng_{\mu\nu}\; d\theta^\mu\otimes d\theta^\nu,
\label{tensor}
\end{equation*}
where $g_{\mu\nu}$ is given by the Eq.\eqref{gFR}.
\begin{remark}
\label{rem1}
The covariance matrix $C$ is a symmetric $n\times n$ matrix. It is well-known that the necessary and sufficient condition for a matrix to be positive is that each of its main minors has strictly positive determinant. So the manifold $\mathcal{M}$ is obtained by requiring that every minors of $C$ has positive determinant.
\end{remark}
Because of Proposition \ref{main} we are able to write the Fisher information matrix as follows
\begin{equation}
G(\theta)=\frac 1 2\left(\begin{array}{cccc}
(c^{-1}_{11})^2&\ldots&\ldots&(c^{-1}_{1n})^2\\
\vdots&(c^{-1}_{22})^2&\ldots&\vdots\\
\vdots&\ldots&\ddots&\vdots\\
(c^{-1}_{1n})^2&\ldots &\ldots&(c^{-1}_{nn})^2
\end{array}\right).
\label{Information}
\end{equation}
Let us consider the inclusion $\iota:\Theta\hookrightarrow \RR^n$; then $\{(\Theta,\iota)\}$ represents an atlas for the manifold $\mathcal{M}$. In this way it straightforwardly follows that it is an orientable Riemannian manifold. Therefore, the Riemannian volume element of $\mathcal{M}$ reads 
\begin{equation}
\nu_g=\sqrt{\det G}\;d\theta^1\wedge\ldots\wedge d\theta^n.
\label{volumelement}
\end{equation}
Recall that $\Theta$ is an open non empty subset of $\RR^n$ and as such is non compact.
Thus, the volume of the manifold is not well defined. 
Let us then introduce the following non negative function
\begin{equation}
\Upsilon(C(\theta)):=\exp\Big[\kappa-
\mbox{Tr}C(\theta)\Big]\log[1+(\det C(\theta))^{n}],
\label{reg}
\end{equation}
where $\kappa\in\RR $ is constant whose role will be explained later on. We observe that this function depends on the network by means of the covariance matrix $C$. The entry $\mu\nu$ of the information matrix \eqref{Information} is given by
\begin{equation}\label{Inversa}
c_{ij}^{-1}=\frac{(-1)^{i+j}\det C[{ji}]}{\det C},
\end{equation} 
where $C[{ji}]$ is the $(n-1)\times(n-1)$ square matrix obtained from $C$ deleting the $j$-th row and the $i$-th column. Moreover we note that the manifold $\mathcal{M}$ is obtained by requiring the covariance matrix $C$ to be positive definite ($ C>0$); as pointed out in Remark \ref{rem1} we have in particular that $\det C>0$. This makes the logarithm in \eqref{reg} meaningful; then, we remark that, from Eq. \eqref{Inversa}, the square root of the determinant of the information matrix $G(\theta)$ in Eq. \eqref{IG} is a ratio of a polynomial in $\theta\in\mathcal{M}$ of degree at most $n-1$ by the $(\det C(\theta))^n$. So, in Eq. \eqref{reg}, we make use of both the negative exponential and the logarithm to avoid $\sqrt{\det G(\theta)}$ blows up to infinity when any direction $\theta_i$ of $\theta$ tends to infinity or when $\det C(\theta)$ approaches to zero. Then the following ``volume" becomes meaningful
\begin{equation}
\mathcal{V}:=\int_{\Theta}\;\Upsilon(C(\theta))\;\nu_g.
\label{volume}
\end{equation}

\begin{remark}\label{rem2}
The quantity \eqref{volume} depends on the intrinsic differentiable structures of $\mathcal{M}$ by means of the volume element \eqref{volumelement} and on the network by means of covariance matrix $C$.
\end{remark}

As we pointed out, for a given $C$ we can identify a well-defined network underlying the statistical Gaussian model. We now give two definitions borrowing notions from graph theory \cite{RD}.

\begin{definition}\label{defiso}
Let $\mathcal{X},\;\mathcal{X}'$ be two networks corresponding to covariance matrices $C$ and $C'$ respectively. We say that $\mathcal{X},\;\mathcal{X}'$ are isomorphic if there exists a permutation matrix $P$ such that
\begin{equation}
C'=P\ C\ P^t.
\label{iso}
\end{equation}
\end{definition} 

It is well-known that a simple and undirected graph has symmetric adjacency matrix, representing which nodes are adjacent to which other ones \cite{RD}. Let us assume that the networks $\mathcal{X},\;\mathcal{X}'$ are simple, undirected graphs with adjacency matrices $A(\mathcal{X}),\ A(\mathcal{X}')$ respectively; it is well-known that two graphs are isomorphic if their adjacencies are congruent via a permutation matrix \cite{RD}. The class of diagonal matrices is invariant under the congruence via permutation matrix. Now, the covariance matrix $C$ of $\mathcal{X}$ is linked to the adjacency matrix in the following way: 
$$C-\left(\begin{array}{cccc}
\theta^1&0&\ldots&0\\
0&\ddots&0&\vdots\\
\vdots&\ldots&\ddots&\vdots\\
0&\ldots&\ldots&\theta^n
\end{array}\right)=A(\mathcal{X}).$$
Thus, from these considerations it straightforwardly follows that if the two graphs are isomorphic then the two networks are isomorphic as well and vice versa.

\begin{definition}\label{defequi}
We say that two networks $\mathcal{X},\;\mathcal{X}'$, corresponding respectively to $C,\ C'$ covariance matrices, are equivalent iff they are isomorphic. In this case we write $\mathcal{X}\sim_{\mbox{iso}}\mathcal{X}'$.
\end{definition}

\begin{pro}
The definition \ref{defequi} is an equivalence relation.
\end{pro}
{\bf Proof.} We have to show that the relation $\sim_{\mbox{iso}}$ is reflexive, symmetric and transitive. First, the relation is reflexive because the identity matrix is a permutation one. Moreover it is symmetric, indeed if $\mathcal{X}\sim_{\mbox{iso}}\mathcal{X}'$ there exists a permutation matrix $P$ such that $C'=PCP^t$. Since $P$ is unitary matrix, we have $C=P^t C' P$ and so $C=QC' Q^t$, where $Q=P^t$. Finally, consider three networks $\mathcal{X},\ \mathcal{X}', \ \mathcal{X}''$, with $C,C',C''$ the corresponding covariance matrices, such that $\mathcal{X}\sim_{\mbox{iso}}\mathcal{X}'$ and $\mathcal{X}'\sim_{\mbox{iso}}\mathcal{X}''$. We know there exist permutation matrices $P$ and $Q$ such that $C'=PCP^t$ and $C''=QC'Q^t$; then $C''=QH C (QH)^t$, where $H=P^t$. So the transitive property follows from the closure under multiplication of the permutation matrices' set. $\hfill\Box$

\begin{definition}
\label{defiF}
Let us consider the set $\mathfrak{X}$ of $n$-vertexes networks modulo the relation $\sim_{\mbox{iso}}$. We define the map ${\mathbf{\mathcal V}}:\mathfrak{X}\rightarrow \RR $ in the following way
\begin{equation}
{\cal V}(C ):=\int_{\Theta}\;\Upsilon(C(\theta))\;\nu_g,
\label{F}
\end{equation}
according to Eq.\eqref{volume} with the specification that the constant $\kappa$ is chosen in order to have ${\cal V}(C )=1$ when $C$ is diagonal.
\end{definition}

We now show that this quantity is well-posed. 

\begin{pro}
The map ${\cal V}:\mathfrak{X}\rightarrow \RR$ defined in \eqref{defiF} is invariant under the equivalence relation 
$\sim_{\mbox{iso}}$.
\label{propF}
\end{pro}

{\bf Proof.} Let $\mathcal{X},\ \mathcal{X}'$ be two networks with covariance matrices $C$ and $C'$, such that $\mathcal{X}\sim_{\mbox{iso}} \mathcal{X}'$. We have to prove that ${\cal V}(C )={\cal V}(C')$.

Consider the function defined in Eq.\eqref{reg}; it is obvious that $\Upsilon(C(\theta))=\Upsilon(C'(\theta))$, indeed the determinant and the trace are invariant under the conjugacy.

Let us now consider the information matrix whose components are given by Eq. \eqref{gFR}. Thanks to Proposition \ref{main}, we have a strict relation between the information matrix and the covariance matrix. Thus we can produce a specific information matrix $G$ (resp. $G'$) for the covariance matrix $C$ (resp. $C'$). We want to prove that $\det G=\det G'$. Because of Eq. \eqref{gvsC} the entries of the information matrix are the square of the entries of the inverse of the covariance matrix (apart from a constant). Since $C'=P C P^t$ then $C'^{-1}=P C^{-1} P^t$, hence it is $\det C'^{-1}=\det C^{-1}$. This is enough to say that $\det G=\det G'$ by applying Proposition \ref{main}.

Finally, we consider the statistical manifolds associated to the networks $\cal X$ and ${\cal X}'$; let us call them $\Theta$ and $\Theta'$ respectively, namely $\Theta=\{\theta| C(\theta)>0\}$ and $\Theta'=\{\theta|C'(\theta)>0\}$. We know that $C'=PCP^t$, so there exists a map $\varphi:\Theta\rightarrow\Theta^\prime$ such that, given $\theta=(\theta^1,\ldots,\theta^n)$, 
\begin{equation*}
\varphi(\theta)=(\theta^{\pi(1)},\ldots,\theta^{\pi(n)}),
\end{equation*}
where $\pi$ is a permutation. It straightforwardly follows that $\varphi$ is a diffeomorphism and its Jacobian 
is such that $\det\mbox{Jac}(\varphi)=\pm 1$. Hence
\begin{eqnarray*}
\int_{\Theta^\prime}f(C^\prime(\theta))\ \sqrt{\det G^\prime(\theta)}\ d\theta&=&\int_{\Theta}f(C^\prime(\varphi(\theta)))\ \sqrt{\det G^\prime(\varphi(\theta))}\ |\det\mbox{Jac}(\varphi)| d\theta\\
\\
&=&\int_{\Theta}f(C)\ \sqrt{\det G}\ d\theta ,
\end{eqnarray*}
being $\Upsilon(C^\prime)=\Upsilon(C)$ and $\det G=\det G^\prime$.
Then we arrive at ${\cal V}(C)={\cal V}(C')$. $\hfill\Box$

\bigskip

We are now ready to define a statistical measure of complexity to be used to state how complex is a network depending on the connections between the nodes (real stochastic variables) it has. 

\begin{definition}
The complexity measure of a network $\cal X$ corresponding to a covariance matrix $C$ of a Gaussian statistical model is given by  
\begin{equation}
{\cal S}:=-\log {\cal V}(C ),
\label{entropy}
\end{equation}
with ${\cal V}(C )$ given by Eq.\eqref{F}.
\end{definition}

\begin{remark}
This definition is inspired by the microcanonical definition of entropy $S$, that is $S:=k_B \log \Omega(E)$, where $\Omega(E)$ is the phase space volume bounded by the hypersurface of constant energy $E$. After integration on the momenta one finds $S = k_B \log \{ \varpi \int_{M_E} [E - V(q_1,\ldots,q_N)]^{N/2} dq_1\dots dq_N\}$, where $ {\varpi}$ is a constant stemming from the integration on the momenta, $M_E$ is the configuration space subset bounded by the equipotential level set $E=V(q_1,\ldots,q_N)$, and $q_1,\ldots, q_N$ are the configurational coordinates. Now, the term $ [E - V(q_1,\ldots,q_N)]^{N/2}$ is just $\sqrt{\det g_J}$, with $g_J$ the Jacobi kinetic energy metric whose associated geodesic flow coincides with the underlying Hamiltonian flow \cite{P07}. In the end the microcanonical entropy is $S=k_B \log \int_{M_E} \sqrt{\det g_J} d\mu(q_1,\ldots,q_N) + k_B \log {\varpi}$, that is proportional to the logarithm of the volume of the Riemannian manifold associated with the underlying dynamics.

Here, by associating a random variable to each node of a network, we can assume that during time
the network - at fixed adjacency matrix - "explores" all its accessible states, so that the
corresponding statistical manifold is the ensemble of all the possible states of the network.
This makes an analogy with ensembles in statistical mechanics. Of course if we define the state 
of a network as a given set of values of the random variables of each node.

\end{remark}


\section{Complexity of simplicial complexes}
\label{section4}

This section is devoted to show the sensitivity of the Entropy in Eq.\eqref{entropy} with respect to the topological features of a network. The latter is now understood as a simplicial complex $K$, which has a purely combinatorial structure. Indeed it consists of a set $\{v\}$ of vertices and a set $\{s\}$ of finite nonempty of $\{v\}$ called simplexes such that
\begin{itemize}
\item[(a)] any set consisting of exactly one vertex is a simplex;
\item[(b)] any nonempty subset of a simplex is a simplex.
\end{itemize}
A simplex $s$ containing exactly $k+1$ vertices is called a $k$-simplex and in such a case the dimension of $s$ is $k$. If $s^\prime\subset s$, then $s^\prime$ is a face of $s$. It follows from condition (a) that the $0$-simplexes of $K$ correspond bijectively to the vertices of $K$. Furthermore, it follows from condition (b) that any simplex is determined by its $0$-faces.

When $K$ is supplied with a topology, the simplexes play the same fundamental role that in linear algebra is played by a basis of a vectorial space. 
As a consequence, the behaviour of a continuous function is determined by its value on the simplexes  \cite{Sp}. Hence, in order to relate the Entropy in Eq.\eqref{entropy} to the topology of the network underlying the Gaussian model of Eq.\eqref{PxT}, it is necessary to understand its behaviour on the simplexes of the network.
Actually, it will be enough to show that the entropy of Eq.\eqref{entropy} varies with the dimension of simplexes to state that it is able to reveal topological features.

To this end, let us start considering the $n$-dimensional Gaussian statistical model of Eq.\eqref{PxT}. We know that the underlying network  (with $n$ nodes) is determined by the covariance matrix $C$. Then, the covariance matrix of a $k-1$ simplex reads
\begin{equation*}
C=\left(\begin{array}{cccc}
C_k&0&\ldots&0\\
0&\theta^{k+1}&\ddots&\vdots\\
\vdots&\ddots&\ddots&0\\
0&\ldots & 0&\theta^n
\end{array}\right),
\label{simplesso}
\end{equation*}
where
\begin{equation*}
C_k=\left(\begin{array}{cccc}
\theta^1&1&\ldots&1\\
1&\theta^2&\ddots&\vdots\\
\vdots&\ddots&\ddots&1\\
1&\ldots & 1&\theta^k
\end{array}\right).
\label{k-simplesso}
\end{equation*}
In order to better understand the information matrix of Eq.\eqref{Information} for a $(k-1)$-simplex, let us write the matrix 
$\Gamma_k$ whose entries are
$\gamma_{k;\,i,j}:=(-1)^{i+j}\det(C_k[i,j])$, where $C_k[i,j]$ is the $(k-1)\times(k-1)$ matrix obtained from $C_k$ by eliminating the $i$th row and the $j$th column. Thanks to the Proposition \ref{main}, in this case the information matrix \eqref{Information} reads
\begin{equation*}
G_k(\theta)=
\frac{1}{2}\left(\begin{array}{cccc}
\frac{1}{(\det C_k)^2}\left(\begin{array}{ccc}
\gamma_{k;\,1,1}^2&\ldots&\gamma_{k;\,1,k}^2\\
\vdots&\ddots&\vdots\\
\gamma_{k;\,1,k}^2&\ldots&\gamma_{k;\,k,k}^2
\end{array}\right)\\
& & \ddots &\\
& & & \Big(\frac{1}{\theta^n}\Big)^2
\end{array}\right).
\label{Informationsimplex}
\end{equation*}
Then it results
\begin{equation}
\det G_k=\Big(\frac{1}{2}\Big)^n\frac{\det \Gamma_k}{\det C_k^{2k}}\prod_{i=k+1}^n\Big(\frac{1}{\theta^i}\Big)^2.\label{DetInformationsimplex}
\end{equation}
Moreover, in this case the function $\Upsilon(C(\theta)) $ of Eq.\eqref{reg} reads 
\begin{equation}
\Upsilon(C(\theta))=\exp\Bigg[\kappa-\mbox{Tr}(C_k(\theta))-\sum_{i=k+i}^n\theta^i\Bigg] \log\Bigg[1+\Bigg(\det C_k(\theta)\prod_{i=k+1}^n\theta^i\Bigg)^n\Bigg],
\label{fsimplex}
\end{equation}
while the parameter space has the following structure
\begin{equation*}
\Theta_k=\Big\{\theta^1>0,\det C_i>0,\theta^j>0\Big\},
\label{thetasimplex}
\end{equation*}
where $i$ and $j$ are integers in $\{2,\ldots,k\}$ and $\{k+1,\ldots,n\}$,respectively. 
Finally, using Eqs.\eqref{DetInformationsimplex} and \eqref{fsimplex} into \eqref{volume} we arrive at
\begin{eqnarray}
\label{volumenk}
\mathcal{V}_k^{(n)}&=&\Big(\frac{1}{2}\Big)^n\int_{\Theta_k}\exp\Bigg[\kappa-\mbox{Tr}(C_k(\theta))-\sum_{i=k+i}^n\theta^i\Bigg] \log\Bigg[1+\Bigg(\det C_k(\theta)\prod_{i=k+1}^n\theta^i\Bigg)^n\Bigg]\nonumber\\
&&\times \frac{\sqrt{\det \Gamma_k}}{(\det C_k)^{k}}\prod_{i=k+1}^n\frac{1}{\theta^i}\;d\theta^1\cdots d\theta^n.
\end{eqnarray}

\bigskip

We are now ready to prove the following result.


\begin{pro}\label{simplex}
Let us consider two $2$-dimensional Gaussian statistical models such that the underlying networks $\mathcal{X}',\mathcal{X}"$  are characterized by the following covariance matrices
\begin{equation*}
C'=\left(\begin{array}{cc}
\theta^1&0  \\
 0 &\theta^2 
\end{array}\right) \qquad
C''=\left(\begin{array}{cc}
\theta^1&1\\
1&\theta^2
\end{array}\right),
\end{equation*}
then we have 
\begin{equation}
\mathcal{V}_1^{(2)}>\mathcal{V}_2^{(2)}.
\label{diseqsimplex}
\end{equation}
\end{pro}

{\bf Proof.}  By Eq. \eqref{volumenk} we have
\begin{equation*}\label{Vn1}
\mathcal{V}_1^{(2)}=\int_{\Theta'} \exp\Big[\kappa-\theta^1-\theta^2\Big]\log\Big[1+\Big(\theta^1\theta^2\Big)^2\Big]\ \Big(\frac{1}{2}\Big)^{\frac{1}{2}}d\theta^1 d\theta^2,
\end{equation*}
and
\begin{equation*}\label{Vn2}
\mathcal{V}_2^{(2)}=\int_{\Theta''}\exp\Big[\kappa-\theta^1-\theta^2\Big]\log\Big[1+\Big(\theta^1\theta^2-1\Big)^2\Big]\Big(\frac{1}{2}\Big)^{\frac{1}{2}}\frac{\sqrt{\big(\theta^1\theta ^2\big)^2-1}}{(\theta^1\theta^2-1)^2}d\theta^1 d\theta^2 \ .
\end{equation*}
We now perform the following change of coordinates $(\theta^1,\theta^2)\mapsto(\theta^1,y/\theta^1)$, which results invertible by recalling that from Eq.\eqref{parameterspace} and Remark \ref{rem1} it is $\theta^1>0$. Then, the above integrals become
\begin{equation*}
\mathcal{V}^{(2)}_1=\int_0^\infty\Bigg(\int_0^\infty \frac{1}{\theta^1} \exp\Big[-\theta^1-\frac{y}{\theta^1}\Big]\ d\theta^1\Bigg)\frac{\log(1+y^2)}y\ dy,\\
\end{equation*}
and
\begin{equation*}
\mathcal{V}^{(2)}_2=\int_1^\infty\Bigg(\int_0^\infty \frac{1}{\theta^1} \exp\Big[-\theta^1-\frac{ y}{\theta^1}\Big]\ d\theta^1\Bigg) \frac{\sqrt{y^2-1}}{(y-1)^2}\log[1+(y-1)^2]\ dy\ .\\
\end{equation*}
Considering further that
\begin{equation*}
\int_0^\infty \frac{1}{\theta^1} \exp\Big[-\theta^1-\frac{y}{\theta^1}\Big]\ d\theta^1=2\  K_0(2\sqrt{y}),
\label{bessel}
\end{equation*}
where the function $K_0( y)$ is the modified Bessel function of the second kind, we arrive
\begin{eqnarray*}
&&\mathcal{V}^{(2)}_1=\int_0^\infty\ 2K_0(2\sqrt{y}) \frac{\log(1+y^2)}{y}\ dy\\
\end{eqnarray*}
and
$$
\mathcal{V}^{(2)}_2=\int_0^\infty 2K_0(2\sqrt{y+1}) \sqrt{1+\frac{2}{y}}\ \frac{\log(1+y^2)}{y} dy.
$$

Now, setting
\begin{equation}\label{}
\varphi(y):=\frac{\log(1+y^2)}{y}\Big(K_0(2\sqrt{y})-\sqrt{1+\frac{2}{y}}K_0(2\sqrt{y+1})\Big),
\end{equation}
to get Eq. \eqref{diseqsimplex} we have to prove that
\begin{equation}
\label{diseq}
\int_0^\infty\ \varphi(y)\ dy >0\ .
\end{equation}
We may notice that $\lim_{y\to 0}\varphi(y)=\lim_{y\rightarrow\infty}\varphi(y)=0$ and
$\varphi^\prime(y)=0$ for $y=y_0$, $0<y_0\ll 1$. Furthermore, $\varphi(y)$ is positive for all $y > y_0$, while it is negative for all $0<y<y_0$. 
Then, Eq. \eqref{diseq} can be rewritten as
\begin{eqnarray*}
\int_{y_0}^\infty\ \varphi(y)\ dy\ >\int_0^{y_0}\ -\varphi(y)\ dy\ .
\label{ineqprincipal}
\end{eqnarray*}
By using the properties of modified Bessel functions of the second kind \cite{W}, we can bound $\varphi(y)$ in the following way
\begin{eqnarray*}
\varphi(y)&\ge& -\frac{K_0(2)}{1+y^2},\quad y\in [0,y_0],\\
\varphi(y)&\ge& \varphi(1)\exp\Big[1-y\Big], \quad y\in [1,+\infty).
\end{eqnarray*}
Thus, we can write the following chain of inequalities
\begin{equation*}\label{sequenza}
\int_{y_0}^\infty\ \varphi(y)\ dy\ > \int_{1}^\infty \varphi(1)\exp\Big[1-y\Big] dy\ > 
\int_0^{y_0} \frac{K_0(2)}{1+y^2} dy
 >\int_0^{y_0}\ -\varphi(y)\ dy\ ,
\end{equation*}
which holds true for $x_0\ll 1$. $\hfill\Box$

\begin{cor}
Under the same hypothesis of Proposition \ref{simplex}, we have 
\begin{equation*}
{\cal S}(C^\prime)<{\cal S}(C^{\prime\prime})
\end{equation*}
\end{cor}
{\bf Proof.} It follows immediately from definition \eqref{entropy}.\hfill $\Box$

\bigskip

\begin{remark}\label{thelast}
Consider the inequality \eqref{diseqsimplex} for $n=3$; that is $\mathcal{V}_1^{(3)}>\mathcal{V}_2^{(3)}$. Setting $K(y):=\Big(K_0(2\sqrt{y})-\sqrt{1+\frac{2}{y}}K_0(2\sqrt{y+1}\Big)$, we should prove that
\begin{equation}\label{caso3}
\int_0^\infty d\theta^3 \Bigg(\int_0^\infty\ \frac{\log(1+(\theta^3 y)^{3})}{\theta^3y}\ K(y)\  dy\Bigg)\exp\Big[-\theta^3\Big]>0\ .
\end{equation}

It would be enough to prove that  the integrand 
of the outer integral is positive, for any value of $\theta^3\in\mathbb{R}_+$.
This amounts to show that  
\begin{equation}\label{case31}
-\int_0^{y_0}\ \frac{\log(1+(\theta^3 y)^{3})}{\theta^3 y}\ K(y)\ dy<\int_{y_0}^{\infty}\ \frac{\log(1+(\theta^3 y)^{3})}{\theta^3 y}\ K(y)\ dy,
\end{equation}
again for any value of $\theta^3\in\mathbb{R}_+$.
Unfortunately we may notice that for $\theta^3\to \infty$ the quantity  $\frac{\log(1+(\theta^3 y)^{3})}{\theta^3 y} K(y)$ is significantly greater than zero only when $y\to 0$ which makes the left-hand side of Eq.\eqref{case31} dominant over the right-hand side.

Nevertheless the result of Eq.\eqref{caso3} is safely obtained by performing the further integration over the variable $\theta^3$, where the cases violating the desired condition \eqref{case31} are suppressed by the factor $\exp[-\theta^3]$. 
\end{remark}

The take home message is that when we try to generalize the relation \eqref{diseqsimplex} to ${\cal V}^{(n)}_k$ with $k<n\in\mathbb{N}$ and $n>2$, we should iteratively consider all integrals over variables $\theta^i$, $1<i \le n$, which appears a task not affordable from an analytical standpoint.
Nevertheless the result of Proposition \ref{simplex} and Remark \ref{thelast} suggest the following conjecture.
\begin{conj}\label{conjec}
Given $n\in\mathbb{N}$, it is 
\begin{equation}
{\cal V}^{(n)}_k>{\cal V}^{(n)}_{k+1},
\end{equation}
for any $0\le k\le n-1$.
\end{conj}
This conjecture is supported by numerical investigations. In Table \ref{table:volume} 
we provide, as an example, Volume \eqref{F} and Entropy \eqref{entropy} vs the dimension of the simplexes
for $n=6$.


 \begin{table}[ht]
\caption{Volume $\cal V$ and entropy $\cal S$ of $k$-simplexes for $n=6$.} 
 \vspace{0.5cm}
\centering
\begin{tabular}{c c c c c}
\hline\hline
$k$ & \hspace{0.2cm} & $\cal V$ & \hspace{0.3cm} & $\cal S$\\ [0.5ex]
\hline
0&  &1 & & 0\\
1& &0.6700 & & 0.5777  \\
2& &0.4024 & & 1.3066 \\
3 & &0.2229  & & 2.1649 \\ 
4 & &0.1158 & & 3.1092 \\
5 & &0.0592& & 4.0767 \\ [1ex]
\hline
\end{tabular}
\label{table:volume}
\end{table}


\section{Concluding Remarks}

The result summarized in the Conjecture \ref{conjec} can be ascribed to the fact that by increasing the dimension of the simplexes, the reduction of the domain of the manifold $\Theta$ (due to the constraints imposed by $C>0$) prevails upon the volume increment due to the change of metric (\ref{Information}), (\ref{volumelement}).

In principle the presented approach can be applied as well to any other multivariate distribution (either continuous or discrete) \cite{And}. 
However, by changing the type of random variables located on the nodes of a network, our quantification of complexity would change. In fact, the geometry of a statistical manifold depends - through the Fisher-Rao metric - on the distribution functions of the random variables defined on the nodes of the network and thus of their type.

Probably the most straightforward extension would be to $t$-distribution and Wishart distribution, being these parametrized by the covariance matrix likewise the normal distribution \cite{And}. 
Thus the regularization introduced in (\ref{reg}) could be also suitable for them. 
More generally it should be adapted to the type of statistical manifold arising from the considered probability distribution function. 
However we would expect that  results similar to the Conjecture \ref{conjec} hold true, provided that the metric changes smoothly with the dimension of simplexes (or whenever the change increases the curvature of the manifold rather than decreasing it).
In any case we might argue that  our measure of networks complexity
not only quantifies complexity in terms of the edges between nodes, but also in terms of type of
random variables located on the nodes.

The topological aspects are not the only possible objects of investigation; there are also combinatorial properties of the networks to take into account. In this work, we mainly focused our attention on the topological ones; but, in order to have a good definition, we showed that if two networks are isomorphic (in the sense of graph theory) the complexity measure introduced does not change its value on them. 
Other combinatorial aspects to address in future investigations would be the behavior of the complexity measure under edge delation/addition, the identification of equivalences classes of graphs with respect to such a measure, etc.

Finally, going far away, we could also envisage the application of our 'static' measure of complexity to time dependent networks  \cite{CNet1} by accommodating time-varying weights on the edges. 
This amounts to considering a time sequence of  
adjacency matrices and, consequently, a time sequence of different statistical manifolds. Thus
we can follow the time evolution of a  network complexity through the time evolution of the volumes 
of the associated manifolds.

\section*{Acknowledgments}
We acknowledge the financial support of the European Commission by the FET-Open grant agreement TOPDRIM, number FP7-ICT-318121. D.F. also thanks Fulvio Lazzeri and Rick Quax for useful discussions.

\end{document}